\documentclass{emulateapj}

\shorttitle{Topological Distance Measures in the Horizon Run 3}
\shortauthors{Speare et al.}

\begin{document}


\title{Horizon Run 3: Topology as a Standard Ruler}


\author{Robert Speare\altaffilmark{1}, J. Richard Gott\altaffilmark{2}, Juhan Kim\altaffilmark{3}, and
Changbom Park\altaffilmark{4}}

\altaffiltext{1}{New York University Abu Dhabi, PO Box 129188, Abu Dhabi, UAE, robert.speare@nyu.edu}
\altaffiltext{2}{Department of Astrophysical Sciences, Peyton Hall, Princeton University,
 Princeton, NJ 08544-1001, USA}
\altaffiltext{3}{Center for Advanced Computation, Korea Institute for Advanced Study,
Heogiro 85, Seoul 130-722, Korea}
\altaffiltext{4}{School of Physics, Korea Institute for Advanced Study, Heogiro 85, Seoul 130-722, Korea}

%\author{Robert Speare}
%\affil{New York University Abu Dhabi, PO Box 129188, Abu Dhabi, UAE}
%\email{robert.speare@nyu.edu}
%\author{J. Richard Gott}
%\affil{Department of Astrophysical Sciences, Peyton Hall, Princeton University,
% Princeton, NJ 08544-1001, USA}
%\author{Juhan Kim}
%\affil{Center for Advanced Computation, Korea Institute for Advanced Study, 
%Heogiro 85, Seoul 130-722, Korea}
%\email{corresponding author: kjhan@kias.re.kr}
%\author{Changbom Park}
%\affil{School of Physics, Korea Institute for Advanced Study, 
%Heogiro 85, Seoul 130-722, Korea}



\begin{abstract}
%We study the Physically Self Bound (PSB) Cold Dark Matter Halo
%distribution which we associate with LRG Galaxies within the Horizon Run 3 and compare the
%topology to that of a Gaussian Random Phase Field (GRF). We apply the routine
%``Contour 3D'' to 108 Mock Surveys of $\pi$ steradians out to
%redshift $z=0.6$ and find that given three separate smoothing lengths
%$\lambda=15,21,34 h^{-1}{\rm Mpc}$, the least $\chi^2$ fit genus 
%per unit volume $g$  yields a $1.7\%$ fractional uncertainty in smoothing
%length and angular diameter distance to $z=0.6$. This is an improvement
%upon former calibrations of and presents a competitive 
%error estimate with next BAO scale techniques.

We study the Physically Self Bound Cold Dark Matter Halo distribution which 
we associate with the massive galaxies within the Horizon Run 3 to estimate the 
accuracy in determination of the cosmological distance scale measured by the topology 
analysis.  
We apply the routine ``Contour 3D'' to 108 Mock Survey of $\pi$ 
steradians out to redshift z = 0.6, which effectively correspond to 
the SDSS-III BOSS survey, and 
compare the topology with that of a Gaussian Random Phase Field.  
We find that given three separate smoothing lengths $\lambda =$ 15, 21, and 34 $h^{-1}{\rm Mpc}$, 
the least $\chi^2$ fit genus per unit volume g yields a 1.7 \% fractional uncertainty 
in smoothing length and angular diameter distance to $z = 0.6$. 
This is an improvement upon 
former calibrations of and presents a competitive error estimate with next
BAO scale techniques.  
We also present three dimensional graphics of the Horizon Run 3 spherical
mock survey 
to show a wealth of large-scale structures of the universe that are predicted in surveys 
like BOSS.
\end{abstract}


\keywords{large-scale structure of the universe -- cosmology:
numerical}

\newpage 

\section{Introduction}
The most popular model for the generation of primordial density
fluctuations is the inflationary scenarios
\citep{guth 1981, linde 1982, albrecht steinhardt, linde 1983}.
This model assumes primordial density perturbations of Gaussian
random phase and it has been shown that such initial conditions
produce a sponge-like topology on large scales \citep[1987]{gott 1986}.
At such scales, where the power spectrum has not been transformed 
by nonlinear growth, the topology of structure in the early universe is well preserved,
and small deviations from random phase predictions give important information
about primordial non-gaussianity, biased galaxy formation, and non-linear clustering \citep{matsubara 1994,park 2005a, park 2012, park gott 1991,park 1998}. 

The genus statistic is central to these studies, and is
now a well-tested quantitative measure 
\citep{gott 1986,gott 1987,hamilton 1986,gott 1989, vogeley 1994, hikage 2002,
hikage 2003, choi 2010,
park 2005a, park 2005b}, having been applied to both the SDSS LRG sample
\citep{gott 2009, strauss 2002, SDSS}, and the CMB \citep{park 1998}. It's utility lies in the
existence of the ``genus curve'', an analytical expression for genus
as a function of density, which allows
comparison of observed topology with that expected from
a standard big bang inflationary model \citep{hamilton 1986}. 

So far, fitting the Gaussian random phase (hereafter, GRP) genus curve to mock surveys
in a $\Lambda$CDM cosmology has been remarkably
successful. The genus has now
been suggested as a cosmic standard ruler \citep{park & kim 2010} and
 a means for probing dark energy \citep{park & kim 2010, zunckel,slepian 2013}. The Baryon
Acoustic Oscillation (BAO) feature, detectable in the power spectrum and galaxy
two point correlation function, is the established ``standard ruler'' \citep{Anderson},
with a reported fractional uncertainty in angular diameter distance to $z=0.6$ of 1.1 \%
expected for the SLOAN survey when completed.
 Now, with the introduction
of ever larger galaxy samples, such as the CMASS Data Release 10 sample of
the SDSS-III Baryon Oscillation Spectroscopic Survey (BOSS), topology is becoming
another attractive technique for probing the expansion of the universe and 
constraining the equation of state of Dark Energy. 
We apply the genus to 108 LRG mock surveys, derived from the Horizon Run 3
$N$-body Simulations \citep{horizonrun}, in order to ascertain the statistical accuracy of 
said ``topological distance measure''. 

\section{The Genus Statistic}
\citet{gott 1986} presented the genus as a reliable description of topology. 
Traditionally, the genus comes from the Gauss-Bonnet theorem, which states that
%the integral of gaussian curvature $K=\frac{1}{r_1r_2}$ (where $r_1$ and 
the integral of gaussian curvature $K=1/({r_1r_2})$ (where $r_1$ and 
$r_2$ are the principle radii) over a compact two-dimensional surface is given
by
\begin{equation}
\int KdA = 4\pi(1-G_b).
\end{equation}
We use a slightly altered form of the Gauss-Bonnet genus, $G=G_b-1$, so
that it has a more intuitive meaning for Cosmology
\begin{equation}
G = (\rm{\#  of\ doughnut\ holes})-(\rm{\# of\ isolated\ regions}). %fix
\end{equation}
See \citet{park 2013} for relation to the Euler characteristic and the Betti numbers. With this definition, the genus of a sphere is $G=-1$; a toroid, $G=0$; three
isolated spheres, $G=-3$; a figure 8 pretzel, $G=1$ (two holes, one isolated body). 
Essentially, the genus is a measure of connectivity. A highly
connected structure -- such as a sponge -- will have many holes, a single body, and
therefore a large, positive genus. A sparse array of objects -- a meatball
topology \citep{soneira peebles,press} -- will have many isolated regions, relatively few holes, and therefore a 
negative genus. An array of isolated voids will also produce a negative genus.

To calculate the genus we smooth the Horizon Run 3 Physically Self Bound Subhalo
distribution \cite{horizonrun} with a Gaussian smoothing ball of radius $\lambda$ 
(Eq. \ref{smoothingball}). 

 We picked the most massive physically bound subhalos to match the number
density of LRG galaxies projected for the SLOAN III survey when completed. The Horizon
Run 3 is a Cold Dark Matter simulation. We make the simple assumption that the most
luminous red galaxies  will from in the centers of the most massive cold dark matter
halos. In the simulation, the most massive subhalos ($>$ 30 CDM particles) are identified
that physically bond and not tidally disruptable by larger structures -- these
we associate with LRG galaxies. 

We then create iso-density contour surfaces of the smoothed density distribution, labeling them
by $\nu$, which is related to the volume fraction $f$ on the high density side of
the contour by
\begin{equation}
f = \frac{1}{\sqrt{2\pi}}\int_\nu^{\infty}e^{-x^2/2}dx,
\end{equation}
Where $x$ is the density parameter.
The value $f=50\%$ corresponds to the median volume fraction contour ($\nu=0$). For GRP
 initial conditions the genus curve is
\begin{equation}\label{genuscurve}
g_{rf}(\nu)=A(1-\nu^2)e^{-x^2/2}.
\end{equation}
%Where the amplitude $A=(\frac{\left<k^2\right>/3)^{3/2}}{2\pi^2})$, and
Where the amplitude $A=(1/2\pi^2) (\left<k^2\right>/3)^{3/2}$, and
$\left<k^2 \right>$ is the average 
value of the squared wave vector, $k^2$ in the smoothed power spectrum \citep{gott 1986};
 or, the slope of the two-point correlation function. 

The shape of a genus curve, and its deviation from the random phase prediction,
can be parametrized by several variables. First, there is the $\chi^2$ best
fit amplitude, which is measured by fitting the GRP curve (
Eq. \ref{genuscurve}) to the observed curve. This gives information
about the power spectrum and phase correlation of the density fluctuation. Secondly,
there are three variables which characterize deviations from a GRF \citep{park 1992}:

\begin{eqnarray}
\Delta \nu &=& \frac{\int_{-1}^{1}g(\nu)\nu d\nu}{\int_{-1}^{1}g_{\rm{rf}}(\nu)\nu d\nu},\\
A_V &=& \frac{\int_{-2.2}^{-1.2}g(\nu)d\nu}{\int_{-2.2}^{-1.2}g_{\rm{rf}}(\nu)\nu d\nu}, \\
A_C &=& \frac{\int_{1.2}^{2.2}g(\nu)d\nu}{\int_{1.2}^{2.2}g_{\rm{rf}}(\nu)\nu d\nu},
\end{eqnarray}
where $g_{\rm{rf}}$ is the best-fit random phase curve (Eq. \ref{genuscurve}). 
$\Delta \nu$ measures any shift in the central part of the genus curve. The 
GRP curve has $\Delta \nu = 0$. A negative value of $\Delta \nu$ is called
a meatball shift, caused by a greater prominence of isolated high-density structures,
pushing the genus curve to the left. $A_V$ and $A_C$ measure the relative 
number of voids and clusters with respect to GRP expectations.

\section{The $N$-body Simulations}
The Horizon Runs, provided by the Korean Institute of Advanced Study (KIAS),
provide some of the best raw material for calibrating topological
study of LRG surveys
\citep{park 2005a,horizonrun}. 
These $N$-body simulations replicate the topology of the
SDSS LRG's exquisitely \citep{gott 2009,lrg sample}.
We use the Horizon Run 3 (HR3) dataset exclusively, which adopts a pressureless
cold dark matter cosmology with a pure cosmological constant $w_\Lambda=-1$.
The basic HR3 cosmological parameters 
were fixed by the WMAP5 data \citep[]{spergel 2003, komatsu 2011, hinshaw 2013}
 and the initial linear power spectrum
was calculated with the CAMB source code \citep{CAMB}. The entire 
simulation is a cube of 374 billion particles, spanning a volume of 
$(10.815 {~ h^{-1}} {\rm{Gpc}})^3$.\footnote{For comparison, this volume is 8800 times larger than the Millenium Run \citep{millenium}.} Initial redshift was
$z=27$ and $N_{\rm step}=600$ discrete timesteps were taken.

%\begin{deluxetable}{rr}
%\tablecolumns{2}
%\tablewidth{0pc}
%\tablecaption{Horizon Run 3}
%\tablehead{
%\colhead{Parameter} & \colhead{Value}
%}
%\startdata
% \omega_m & .26 \\
%\enddata
%\end{deluxetable}

\subsection{Mock LRG Survey Construction}

The selection of cold dark matter halos uses the Friend
of Friend algorithm, where separation cut off distance is 20 \% of the mean 
separation distance. To improve cluster identification, HR3 searches for 
Physically Self Bound (PSB) subhalos that are gravitationally self-bound and not tidally disruptable 
\citep{kim and park 2006}.
This provides a substantial increase in the similiarity between simulation and 
observational data, as these dark matter subhalos are sites for LRG formation. 

To simulate the SDSS survey dimensions, HR3 places 27 observers evenly within 
its cubical volume and allows each observer to see out to a redshift of $z < 0.7$. 
This crates 27 independent, non-overlapping spherical regions.
The co-moving positions and velocities of all CDM particles are saved as they cross
their past light cone and PSB subhalos are identified from this data. In preparation
for the SDSS-III LRG catalogue, it was assumed that a volume-limited sample would 
yield a constant number density of $3 \times 10^{-4} (h^{-1}{\rm Mpc})^3$. 
In order to match this prediction, the minimum mass limit of the PSB subhalos was 
varied with redshift and
 the absolute minimums were set to $9.75 \times 10^{12}~{h^{-1}}{\rm M_{\odot}}$. Given
these parameters, the physical properties of the HR3 mock surveys match very well
with the most recent LRG surveys \citep{choi 2010,gott 2009, gott 2008}.

\begin{figure}[tpb]
\epsscale{1.0}
\plotone{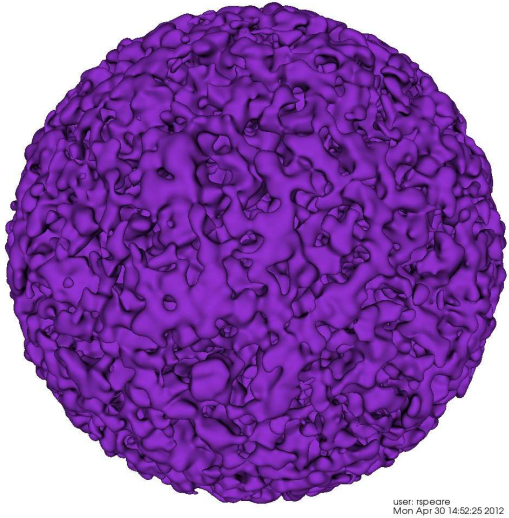}
\caption{A spherical Horizon Run 3 Mock survey out to
redshift $z=0.7$. The PSB subhalo counts have been smoothed with
a Gaussian smoothing ball of $\lambda=34~h^{-1} \rm{Mpc}$. 
See \ref{3Dappendix} for 3D plots of the Horizon 3 data.}\label{HR3mock}
\end{figure}

\section{Methods}
\subsection{Smoothing and Discretization}
We smooth the 27 past lightcone 
PSB subhalo distributions with a gaussian smoothing ball
\begin{equation}\label{smoothingball}
W(\vec{r}) = \frac{1}{(2\pi)^{3/2}}e^{-\frac{r^2}{2\lambda^2}},
\end{equation}
smearing structure on scales smaller than $\lambda$. The Mock survey data
is placed into a three-dimensional pixel grid of density values, and we choose
$\lambda$ to always be greater than $2.5$ pixel sidelengths $s$.
 For cold dark matter models, smoothing with a 
Gaussian recovers the topology of the initial density field, provided
that the smoothing length $\lambda$ is sufficiently greater than the correlation length
$R_0$ and non-linear effects are avoided\footnote{$R_0$ is approximately
$5 ~h^{-1}\rm{Mpc}$ for LRG}. 
\subsection{Conversion and Trimming}
With this smoothed Mock Survey in hand, we convert from co-moving spherical coordinates
to redshift coordinates, using a comoving line of sight distance formula
\citep{hogg}. PSB subhalo peculiar velocities are converted into 
redshift distortions by
\begin{equation}
\Delta z = \frac{\vec{v_{\rm{r}}}}{c} = \frac{\hat{r}\cdot \vec{v}_{\rm{pec}}}{c},\\
\end{equation}
where $\vec{v}_r$ is the radial velocity, $\hat{r}$ is the unit radial vector,
and $\vec{v}_{\rm pec}$ is the cartesian peculiar velocity of the subhalo.
After redshift converting and correcting, we save PSB subhalo counts within
 a grid of dimensions $650^3$, with cubical pixel volume of 
$s^3=(6 {~ h^{-1}}{\rm{Mpc}})^3$. The entire grid spans a volume of $(1950 {~ h^{-1}}{\rm{Mpc}})^3$. 

We then apply an angular mask, splitting the 27 perfectly spherical mock
surveys into four quadrants each of $\pi$ steradians and 
radius $z=0.6$, to approximate the area of sky coverage and depth in the SLOAN III survey.
With these $4\times 27=108$ smoothed mock surveys in hand, we calculate
the genus using a polygonal approximation scheme developed by \citet{weinberg1988,hamilton 1986} 
called ``Contour 3D'', which adds up angle deficits at pixel vertices. 

\section{Using Topology as a Standard Ruler}
\label{protocol}
\begin{figure}[tpb]
   \centering
   \plotone{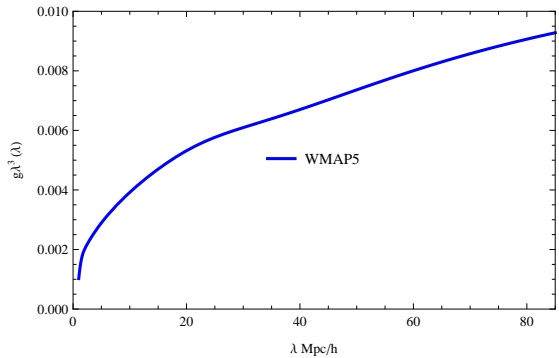}
   \caption{Genus per cubic smoothing length $g\lambda^3$ 
   for the WMAP5 parameters ($\Omega_m=0.26$, $H_0=74$),
   assuming a flat $\Lambda$CDM cosmology \citep[taken from Fig. 1 of][calculated by Young Rae Kim]{zunckel}.}
   \label{glambda3}
\end{figure}

An application of quantitative topology being applied to the SDSS LRG sample --
other than testing the gaussianity of initial density fluctuations -- is
to measure cosmological parameters, such as those governing the expansion
history of the universe. This can be done by measuring the genus
statistic within a fixed volume at different redshifts.
 In the instance of $N$-body simulations, one knows the
 correct cosmological model and therefore the correct transformation $r \to z$.
One smooths the density field with a known
smoothing length $\lambda$ and then measures the median density genus within a 
volume $V$. This yields $g=G/V$, genus per unit volume, which
one can use to indirectly measure any physical volume by counting structures.
In order to more explicitly state the smoothing length dependence, the dimensionless
quantity $g \lambda^3$ is often used, which is simply the genus per cubic smoothing 
length. This quantity can be analytically calculated from a full set of cosmological
parameters and a linear power spectrum. Such a function $g \lambda^3 (\lambda)$ has
been examined closely for the WMAP3 and WMAP5 parameters (see fig. 1 of \citealt{park & kim 2010} and fig. 1 of \citealt{zunckel}, drawn by Y.R. Kim.)(see Fig. \ref{glambda3}).

In practice, we do not know the true cosmological model.
Let us illustrate the effects of applying an incorrect
cosmological model to a survey sample. If we underestimate the expansion
rate of the universe $H_0$, then our conversion from redshift to comoving space
will put celestial objects too far from the Earth. This causes an overestimation
of survey volume. For a homogeneous and isotropic survey, genus is linearly
proportional to volume and therefore an overestimation of $V$ will drive the genus
at a certain smoothing length up ($G(\lambda) \uparrow$). At the same time however,
we have also adopted a co-moving smoothing length $\lambda$ that is larger than
intended. This will change the actual scale of study and erase all structure beneath
scale by convolution, decreasing the genus ($G(\lambda) \downarrow$). Luckily, the
net effect is detectable since the amplitude $G$ of the genus curve effectively measures the slope
of the power spectrum at the scale $\lambda$, which is not scale invariant\citep{park & kim 2010}.

Our procedure for measuring angular diameter co-moving distance to $z=0.6$ is 
straightforward. We assume a $\Lambda$CDM flat cosmological model. 
$\Omega_m$, $h$, and $\Omega_\Lambda$ come from CMB fits with $l > 210$ which are
insensitive to $w_\Lambda$ because dark energy has negligible influence at recombination.
These values are used to construct the power spectrum and from that, $g\lambda^3(\lambda)$ (see Fig. 
\ref{glambda3}). Now we measure $g\lambda^3$ and get a value; we look on our analytical plot -- Fig. \ref{glambda3} --
and find the true value of $\lambda$, which we will call $\lambda_{\rm{true}}$. If this
is 1\% smaller than the initial value of $\lambda$ that one used, it means that the 
co-moving distance out to $z=0.6$ is also 1\% smaller than previously thought. 
In this way one can measure co-moving distance out to $z=0.6$. And, with this 
as one data point one can fit a cosmological model, leaving $w_\Lambda$ as a parameter 
\citep{park & kim 2010}.

If the intial cosmological model is slightly wrong ({\it i.e.}
$w_\Lambda$ may not be exactly $-1$, or may vary with time; \citealt{slepian 2013}), 
this is inconsequential because we are just measuring the topology --
counting the total number of structures inside $z=0.6$. If the
radial co-moving distance inside this volume is proportionately in error it will
make no difference, as that will just distort shapes and structures slightly
without altering their count \citep[see][]{zunckel}. An rms cosmic variance in the total genus
$\sigma_g$ out to $z=0.6$ in a survey sample will cause a fractional
rms error of ${\sigma_g}/{g}$ in $g\lambda^3$; and given the slope of the curve,
$(g\lambda^3)'$ at the applied $\lambda$, this will introduce an rms error
in $\lambda$ and therefore in co-moving distance at $z=0.6$ of:

\begin{equation}
%\frac{d}{d\lambda}
(g\lambda^3)'\frac{\sigma_\lambda}{\lambda} = \frac{\sigma_{g}}{g}.
\end{equation}

\subsection{Uncertainties in such a ruler}
We examine the statistical variance of genus per unit volume
$g$ in the Horizon Run 3 mock surveys, which is far from an ``ideal'' measurement.

An ``ideal'' measurement of $g$ would be
 examining the initial density field in comoving space. If the initial conditions were of
GRP, one would expect excellent agreement between the observed genus
curve and the theoretical GRP curve; however, finite sample size, even at this level,
 introduces an error because of no power at large
scales (or, larger than the simulation box size). The next best measurement of $g$ would be examining
 the final conditions of the entire $N$-body simulation in comoving space, which erases a 
portion of the cosmic variance associated with small survey size,
 but is subject to the effects of non-linear gravitational
infall and galaxy formation bias. An unavoidable source of error, ``ideal'' or otherwise,
is finite pixel resolution, which applies a smoothing scale to the data and
destroys structure smaller than pixel size $s$.

 Observation of $g$ in comoving
space has obvious advantages to observation in redshift space, since one has
complete knowledge of all PSB subhalo positions and velocities.
It has been found that the redshift correction for peculiar velocity
presents the worst source of error for the $\chi^2$ best fit amplitude of the genus curve
\citep{choi 2010}. The application of peculiar velocity redshift corrections
is in essence a smoothing routine of its own, in that real-space structures
are radially smeared due to ``fingers of god'' effects. This effectually
raises the observed smoothing parameter $\lambda$ slightly and yields a lower
$\chi^2$ best fit genus amplitude. The choice of survey volume, specifically
volume to surface area ratio, also creates error because of data being ``smoothed out''
of the survey region. The complicated boundaries of the SDSS present a cause
for concern; particularly the three thin stripes along the southern Galactic cap,
which are ignored altogether during genus analysis. 

%Observation of the CMB gives us the initial power spectrum out to scales of 
%$144\rm{Mpc}$. 
%Prior to recombination, $\Omega_\Lambda$ and $w_\Lambda$ are unimportant. Scales coming
%inside the Horizon prior to recombination tell us $\Omega_m h^2$ and $P(k)$ out to scales of $108 \frac{\rm{Mpc}}{\rm{h}}$, scales much larger than the smoothing lengths we will
%be considering \citep{zunckel}. 

An SDSS measurement of $g$ uses a finite, redshift space sample, 
where the aforementioned sources of error apply: the cosmic variance 
associated with small survey size;
non-linear clustering; boundary effects; and redshift space distortion.
The situation sounds daunting, but because of its
size, the Horizon Run 3 provides an ensemble of tests.
 We split the 27 HR3 spherical mock surveys into four quadrants, thereby
acquiring 108 ``genus experiments'' for a chosen smoothing length $\lambda$.
 \citet{gott 2009}
have reported the genus amplitude of the SDSS LRG to within 5\% accuracy. Based upon our 
results (see Table \ref{statresults}), we believe that this fractional uncertainty can be reduced to
of 1\%.

\section{Results}
We measured the genus per cubic smoothing length for $\lambda=15$, 21, and 34 $~h^{-1} \rm{Mpc}$,
studying the random and systematic error over 108 HR3 mock surveys. 
 For $\lambda=15 ~ h^{-1}{\rm{Mpc}}$, the fractional uncerainty in 
genus per cubic smoothing length was less than one percent, which translates
 to a fractional uncerainty in smoothing length -- and angular diameter distance -- 
of approximately $2.1$\% (Table \ref{statresults}). 

Treating the variance at $\lambda=15$, 21, and 34 $h^{-1}{\rm{Mpc}}$ as statistically 
independent -- since HR3 adopts a random phase model and the smoothing volumes
are significantly different --
we add the three smoothing length rms errors in quadrature
\begin{equation}
\frac{1}{\sigma_{\rm{eq}}^2} = \frac{1}{\sigma_1^2}+\frac{1}{\sigma_2^2}+\frac{1}{\sigma_3^2},
\end{equation}
 yielding a $1.69$\% fractional uncertainty in smoothing length and angular diameter
distance out to $z=0.6$. Combining only the $21$ and $34~h^{-1}$Mpc samples, we get a 
$2.97$\% 
fractional uncertainty in smoothing length.

\begin{table*}[htb]
\renewcommand{\arraystretch}{1.5}
\begin{center}
\begin{tabular}{|l|l|l|l|}\hline
& $\lambda=15~{\rm{Mpc}}/h$ 
& $\lambda=21~{ \rm{Mpc}}/h$ 
& $\lambda=34 ~{\rm{Mpc}}/h$ \\ \hline
%G  & 5718.42 & 2364.58 & 646.66 \\
$g \lambda^3\times 10^3$  & 4.762 & 5.403 & 6.271 \\
$\sigma_{g \lambda^3}\times 10^3$  & 0.04380 & 0.6732 & 1.358 \\
$\frac{\sigma_{g \lambda^3}}{g \lambda^3}$ & .919\%  & 1.245\% & 2.166\% \\
$\lambda_{\rm{t}}$ & 15.448 & 20.823 & 32.993\\
$\frac{\lambda_{\rm{t}}-\lambda}{\lambda}$ & 2.99\% & -0.84\% & -2.96\% \\
$\frac{\sigma_{\lambda}}{\lambda}$ & 2.096\% & 3.215\% & 6.742\%\\
% $A_V$ & 0.809$\pm$0.011 & 0.860 $\pm$ 0.017 & 0.92 $\pm$ 0.03\\
% $A_C$ & 1.10 $\pm$ 0.015 & 1.08 $\pm$ 0.02 & 1.06 $\pm$ 0.03\\ 
\hline
\end{tabular}
\end{center}
\caption{$g\lambda^3$ is the averaged $\chi^2$ best-fit genus per cubic smoothing
length, for all 108 mock surveys -- multiplied 
by $10^3$ for the reader's sake. $\lambda_{\rm{t}}$ is the corresponding ``true''
smoothing length for the observed genus per cubic smoothing length, as discussed in
Section \ref{protocol}. $\sigma_{g\lambda^3}$ and $\sigma_\lambda$ represent
the variance in smoothing genus per cubic smoothing length and smoothing length. 
$({\lambda_t-\lambda})/{\lambda}$ is the fractional, systematic error in 
smoothing length.} 
\label{statresults}
\end{table*} 

\begin{figure}[tpb]
   \centering
   \plotone{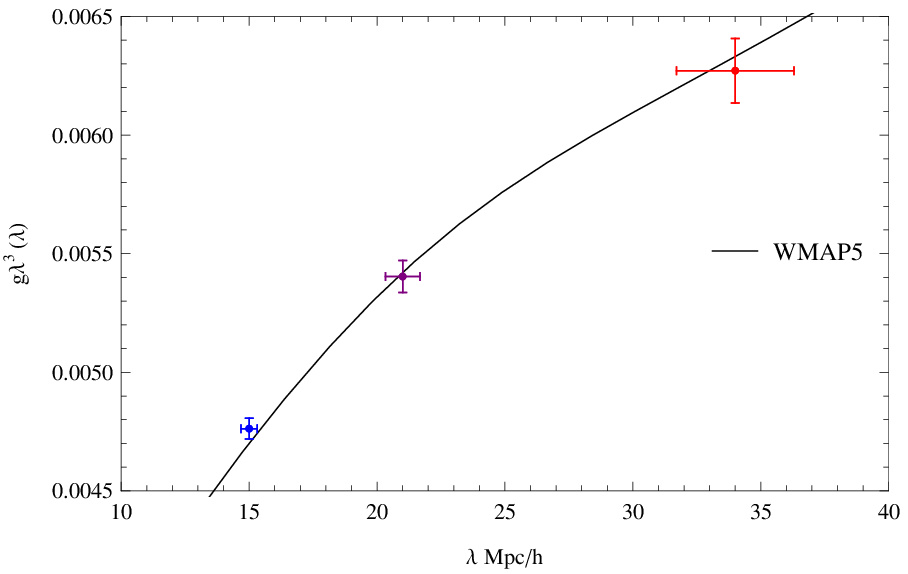}
   \plotone{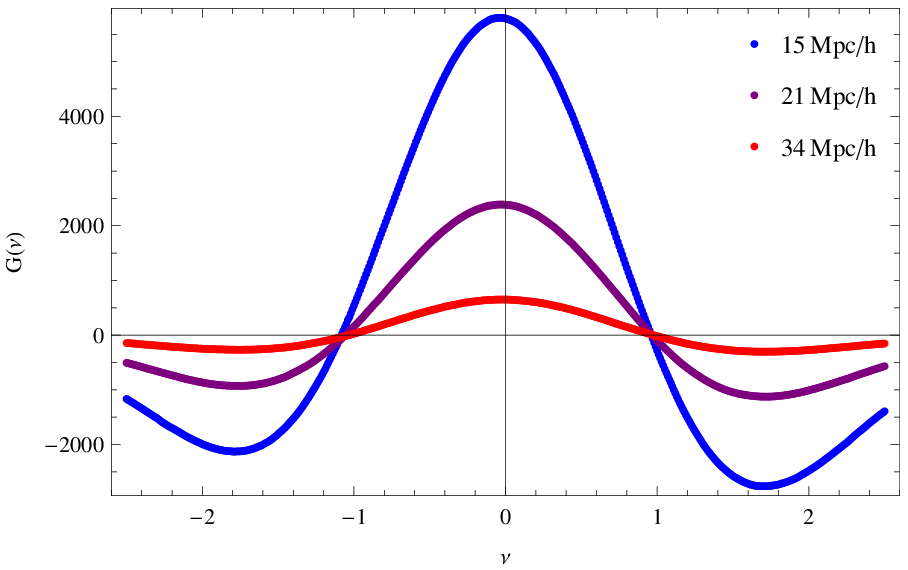}
   \caption{Above: genus per cubic smoothing length $g\lambda^3$ 
   for the WMAP5 parameters, with $\chi^2$ best-fit
   data points and $1\sigma$ error bars.
   Below: the ensemble averaged genus curves $G(\nu)$ for $\lambda=15$, 21, and 34
   $h^{-1} {\rm{Mpc}}$ }
   \label{glambda3withdata}
\end{figure}

With 108 samples in hand, our fractional ``uncertainty of the uncertainty'' is
 $1/\sqrt{2(N-1)}=6.8\%$.  It is notable that the systematic effect
for the $21 ~ h^{-1} $Mpc sample was very small, $-0.84$\%, and that the $\chi^2$ best-fit 
genus amplitudes modeled the $g \lambda^3$ curve extraordinarily well (Fig. 
\ref{glambda3withdata}).

\section{Discussion}

With these results in hand, it is important to continue refining topological 
study of the SDSS LRG sample with $N$-body simulations. Extremely large cubes like HR3
allow for tight description of the cosmic variance in genus per unit volume $g$ and
smoothing length $\lambda$. This statistical knowledge translates directly to the
measurement of cosmological parameters such as $w$. A possible extension of this
work is to more accurately model the SDSS survey with 108 less 
``ideal'' masks. Another possible extension is to measure the variance in genus
per cubic smoothing length $g\lambda^3$ for a large number of $\lambda$'s, perhaps
iterating from $15~h^{-1}{\rm Mpc}$ to $34~ h^{-1}{\rm Mpc}$ in small increments 
$\Delta \lambda < 0.2~h^{-1}{\rm Mpc}$. Smooth plots of $\sigma_{g\lambda^3}$ and 
$\lambda - \lambda_{\rm true}$ as a
function of $\lambda$ could yield useful information about the evolution 
of random and systematic error with scale.

\acknowledgments
We thank the Princeton Department of Astrophysical Sciences, Princeton NJ, 
where this work was completed. We thank the support of the Supercomputing Center/Korea Institute of Science and Technology Information with supercomputing resources, including
technical support (KSC–2011–G2–02) for Horizon Run 3. We also thank Korea Institute for Advanced Study for providing computing resources (KIAS Center for Advanced Computation Linux Cluster System).


%% in parentheses, after the keyword, but they will not be verified.

%{\it Facilities:} \facility{Nickel}, \facility{HST (STIS)}, \facility{CXO (ASIS)}.



\appendix
%\section{Tables}

\section{3D Graphics of the Horizon Run 3}
\label{3Dappendix}

\begin{figure}[tpb]
\epsscale{1.0}
\begin{center}
\plottwo{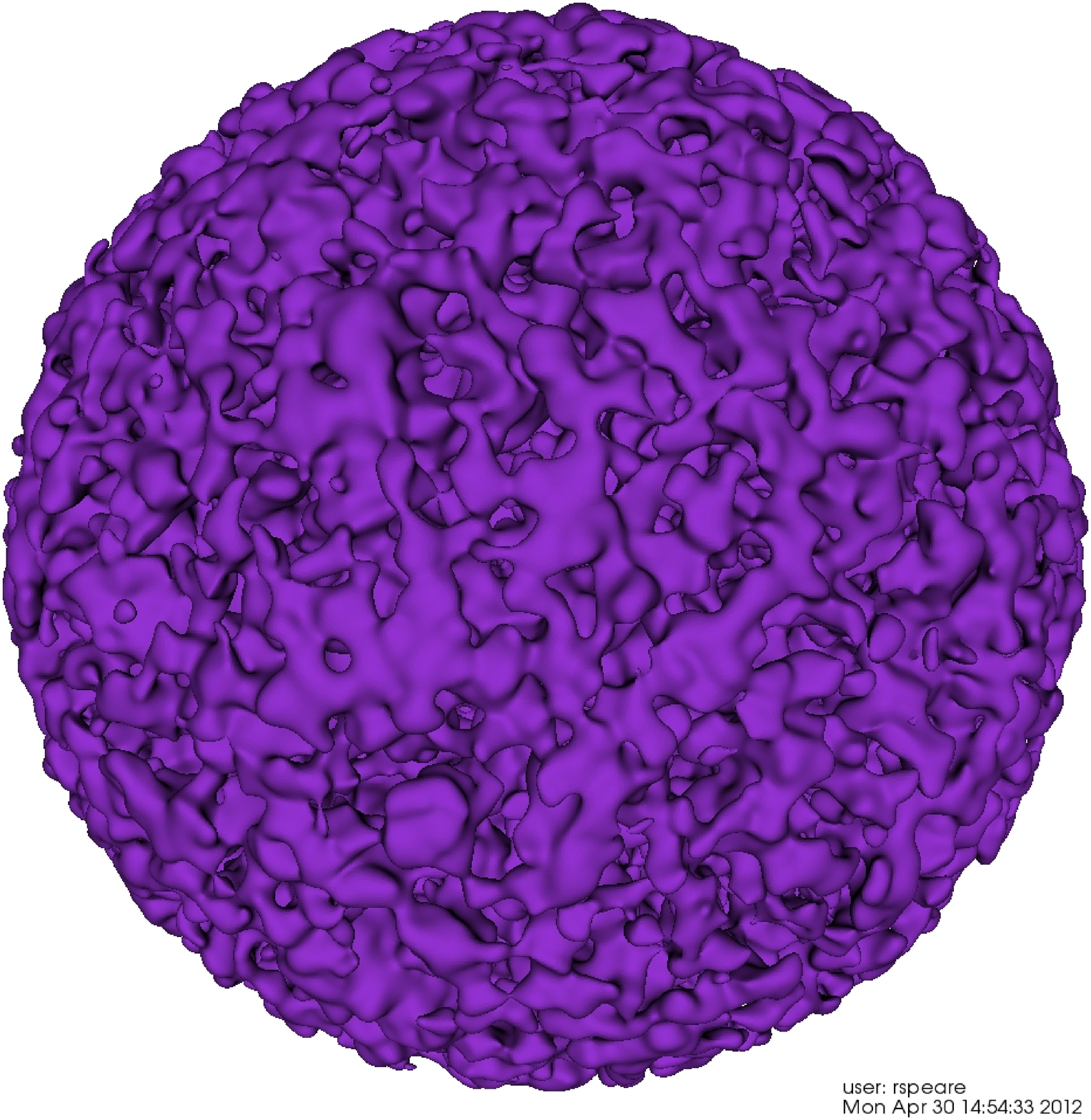}{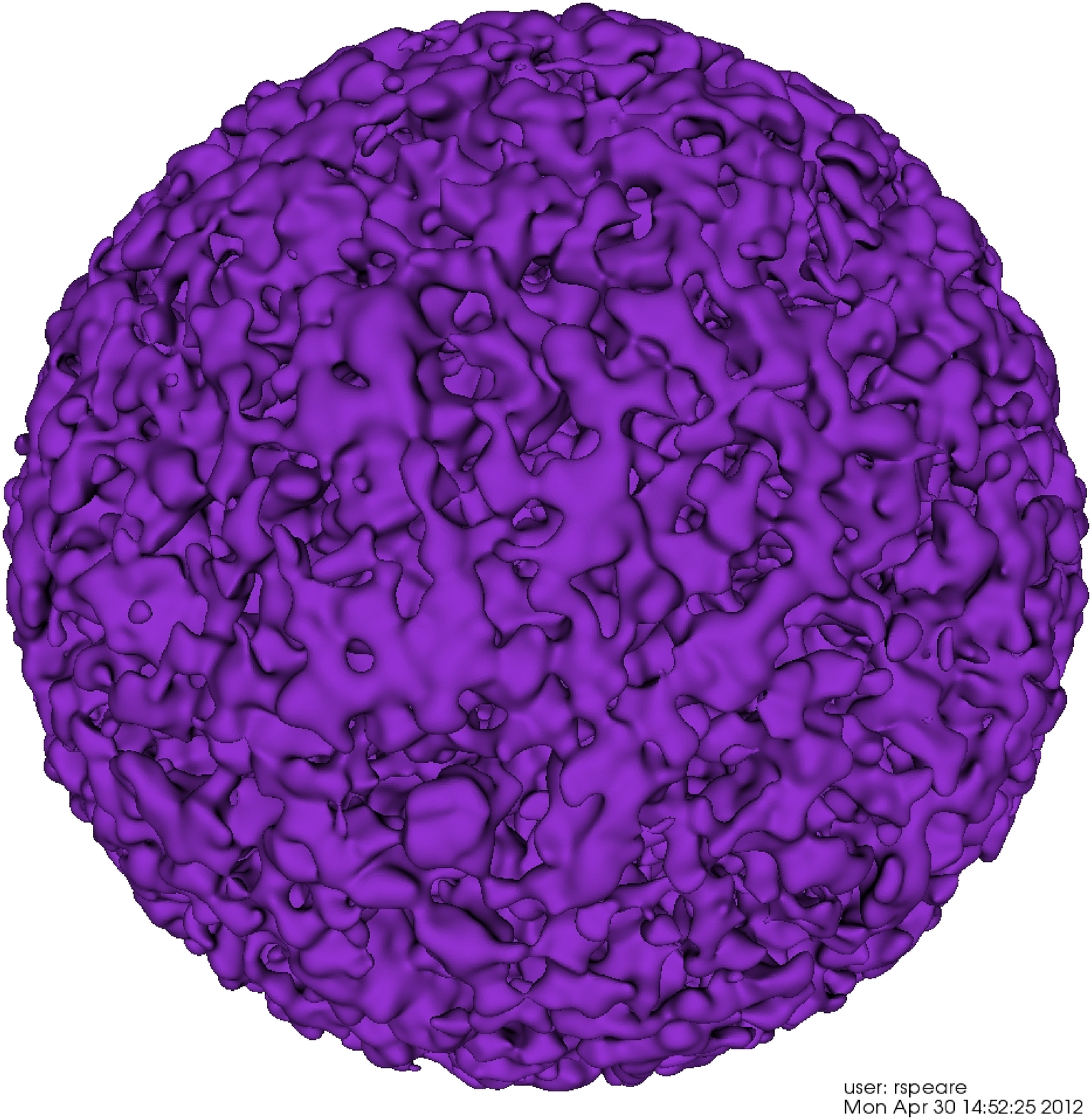}
\caption{Stereo View of the Horizon Run 3 spherical mock survey out to redshift
$z=0.7$. In order to see this figure in 3D, cross you eyes until you see
three purple sponges. Fuse the middle sponge and bring it into focus. 
The middle sponge should be in 3D.}\label{HR3stereoview}
\end{center}
\end{figure}

\begin{figure}[tpb]
\epsscale{1.0}
\plotone{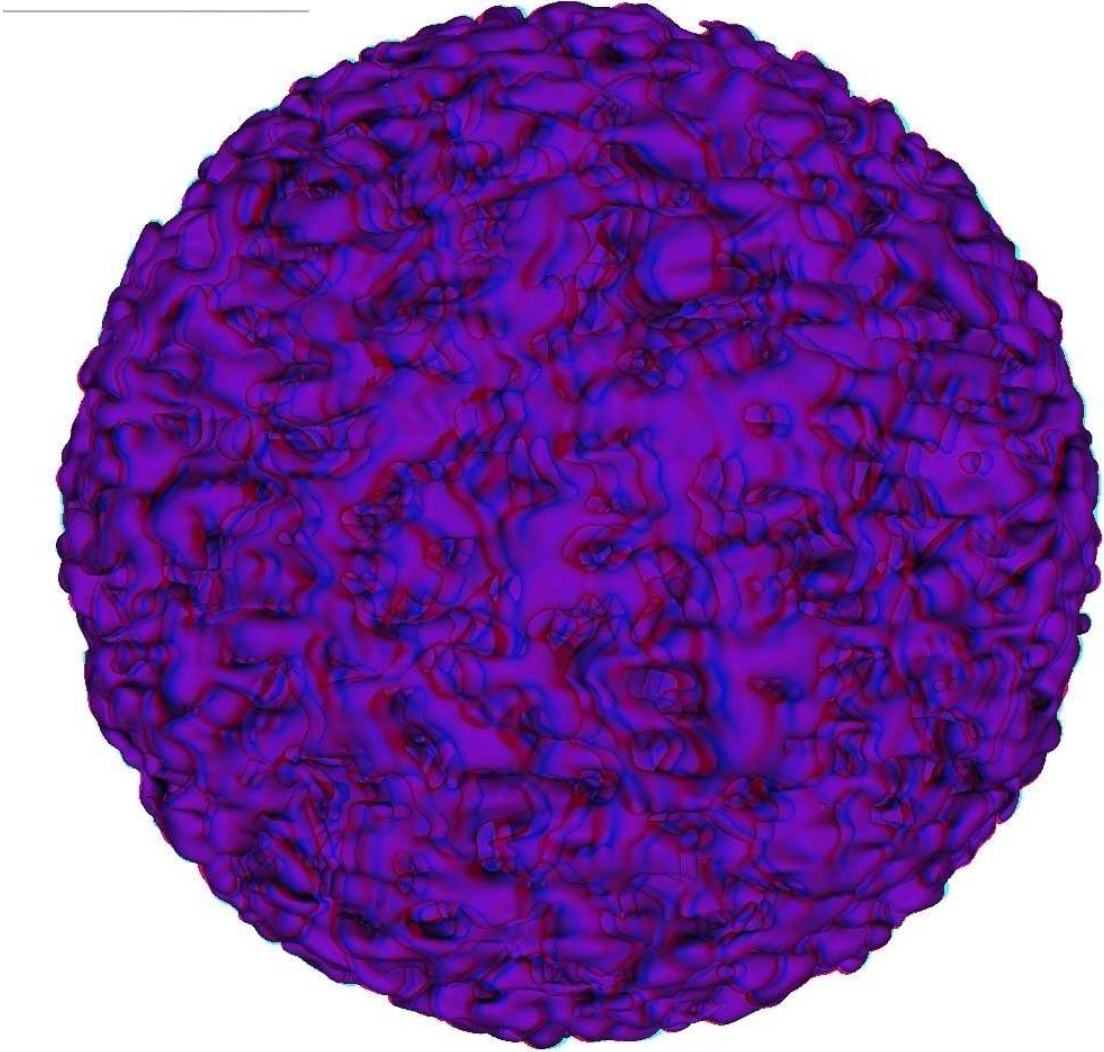}
\caption{A red blue anaglyph of a spherical Horizon Run 3 Mock survey out to
redshift $z=0.7$. The PSB subhalo counts have been smoothed with
a Gaussian smoothing ball of $\lambda=34~h^{-1} \rm{Mpc}$. Requires
red-cyan lensed glasses to view.}\label{HR3anaglyph}
\end{figure}




\clearpage

%% the file, uncomment the \plotone, \plottwo, and
%% If you need a layout that cannot be achieved with \plotone or
%% \plottwo, you can invoke the graphicx package directly with the
%% \includegraphics command or use \plotfiddle. For more information,
%% please see the tutorial on "Using Electronic Art with AASTeX" in the
%% documentation section at the AASTeX Web site, http://aastex.aas.org/
%%
%% The examples below also include sample markup for submission of
%% supplemental electronic materials. As always, be sure to check
%% the instructions to authors for the journal you are submitting to
%% for specific submissions guidelines as they vary from

%% This example uses \plotone to include an EPS file scaled to
%% 80% of its natural size with \epsscale. Its caption
%% has been written to indicate that additional figure parts will be
%% available in the electronic journal.

\clearpage

%% Here we use \plottwo to present two versions of the same figure,
%% one in black and white for print the other in RGB color
%% for online presentation. Note that the caption indicates
%% that a color version of the figure will be available online.
%%

%% This figure uses \includegraphics to scale and rotate the still frame

%% If you are not including electonic art with your submission, you may
%% mark up your captions using the \figcaption command. See the
%% No more than seven \figcaption commands are allowed per page,
%% so if you have more than seven captions, insert a \clearpage

%% Tables should be submitted one per page, so put a \clearpage before

%% Two options are available to the author for producing tables:  the
%% deluxetable environment provided by the AASTeX package or the LaTeX
%% table environment.  Use of deluxetable is preferred.
%%

%% Three table samples follow, two marked up in the deluxetable environment,
%% one marked up as a LaTeX table.

%% In this first example, note that the \tabletypesize{}
%% command has been used to reduce the font size of the table.
%% We also use the \rotate command to rotate the table to
%% landscape orientation since it is very wide even at the
%% Note also that the \label command needs to be placed

%% This table also includes a table comment indicating that the full
%% version will be available in machine-readable format in the electronic

\clearpage

%% Tables may also be prepared as separate files. See the accompanying
%% sample file table.tex for an example of an external table file.
%% To include an external file in your main document, use the \input
%% command. Uncomment the line below to include table.tex in this
%% sample file. (Note that you will need to comment out the \documentclass,
%% \begin{document}, and \end{document} commands from table.tex if you want
%% to include it in this document.)


%% The following command ends your manuscript. LaTeX will ignore any text

\end{document}